\documentclass[pra,showpacs,twocolumn,showkeys]{revtex4-1}
\usepackage{graphics}
\usepackage{graphicx}
\usepackage{amsmath}
\usepackage{color}
\usepackage{bbold}
\usepackage{cancel}

\begin{document}

\title{Geometry-Controlled Nonlinear Optical Response of Quantum Graphs}
\author{Shoresh Shafei}
\author{Rick Lytel}
\author{Mark G. Kuzyk}
\affiliation{Department of Physics and Astronomy, Washington State University, Pullman, Washington  99164-2814}

\begin{abstract}
{\noindent We study for the first time the effect of the geometry of quantum wire networks on their nonlinear optical properties and show that for some geometries, the first hyperpolarizability is largely enhanced and the second hyperpolarizability is always negative or zero. We use a one-electron model with tight transverse confinement.  In the limit of infinite transverse confinement, the transverse wavefunctions drop out of the hyperpolarizabilities, but their residual effects are essential to include in the sum rules.  The effects of geometry are manifested in the projections of the transition moments of each wire segment onto the 2-D lab frame.  Numerical optimization of the geometry of a loop leads to hyperpolarizabilities that rival the best chromophores.  We suggest that a combination of geometry and quantum-confinement effects can lead to systems with ultralarge nonlinear response.}
\end{abstract}

\maketitle

\section{Introduction}
The study of nonlinear optical (NLO) properties of materials has been the subject of extensive research due to the interesting underlying physics as well as the wide application range in science and technology. Concerted efforts by theorists, synthetic chemists, materials physicists, and device designers have focused on optimizing hyperpolarizabilities, the quantities that govern all light-matter interactions, to enable new photonic materials for applications in all-optical switching \cite{hales10.01}, 3-D photolithography \cite{kawat01.01}, quantum information,\cite{feizp11.01} optical data storage \cite{cumps99.01}, photodynamic cancer therapies \cite{roy03.01} and other fields.

In the past decade, the theory of fundamental limits found that the first- and second-order hyperpolarizabilities in the off-resonance regime,\cite{kuzyk00.01,kuzyk00.02} denoted by $\beta$ and $\gamma$ respectively, depend on the number of electrons contributing and the energy difference between the ground and the first excited states of the system.  A comparison of the theory with experiment showed the first hyperpolarizability (which we call the hyperpolarizability throughout this text) of the best molecules fell short of the fundamental limit (FL) by about a factor of 30. Similar results were found for the second hyperpolarizability, $\gamma$.

Monte Carlo simulations of the first \cite{kuzyk08.01} and second \cite{shafe10.01} hyperpolarizabilities confirm these limits.  Based on numerical optimization of a large variety of potential energy functions, the best hyperpolarizabilities are found to be about $0.71$ of the FL.\cite{zhou06.01,zhou07.02} Using a set of experimental and computational tools including linear spectroscopy, Raman spectroscopy, $\beta$ values measured by Hyper-Rayleigh scattering and Stark spectroscopy, Tripathy \emph{et al.} studied possible origins of the gap, including dilution effects due to vibronic states, unfavorable energy spacing, truncation of the sum rules used in the theory of fundamental limits and smaller effects such as simplifications of the Lorentz local field model and systematic errors in experiments. Experiments suggest that the nature of the energy spectrum spacing is the culprit.\cite{Tripa04.01}

The theory of FL predicts that progressively larger energy spacing leads to larger $\beta$ values. However, most molecules have energy spacing that decreases with increasing number of states. Recent Monte Carlo simulations for different classes of energy spectra, including hydrogen-like atoms, a harmonic oscillator and a particle in a box clearly show that energy spacing is the key factor, independent of transition moments.\cite{shafe11.01}  Our studies of nanowires are motivated by the fact that a particle in the box has favorable energy spacing that is not seen in molecules.

Frohlich first pointed out that at the limit of small metal particle size, the continuous electronic conduction band becomes discrete.\cite{frohl37.01} Kubo used the one-electron approximation to show that at low temperatures, the electronic states of such systems are comparable to $kT$.\cite{kubo62.01} Jaklevic \emph{et al.} reported the observation of discrete states in crystalline Pb \cite{jakle71.01}, and Brus' model of semiconductor crystallites predicted the shape and size dependence of the state energies.\cite{brus84.01} Ashoori reviews the field of \emph{artificial} atoms.\cite{ashoo96.01}

Electrons in quantum wires are confined in 2-D. Semiconductor nanowires typically have diameters of 1-100 nm and lengths of several micrometers.  Key parameters, such as chemical composition, diameter, and length can be controlled in their synthesis, enabling a wide range of devices and applications such as p - n diodes, LED's, transistors and nano-scale lasers.\cite{liebe07.01}. For example, Huang \emph{et al.} demonstrated the ultraviolet lasing in ZnO nanowires at room temperature, which can be used as a nano-scale laser source in optical computing and data storage.\cite{huang01.01} Johnson \emph{et al.} used near-field scanning optical microscopy technique to study the NLO properties of ZnO nanowires and suggested their use as frequency converters.\cite{johns02.01} Nanowire structures can also be used in light generation, propagation, detection, amplification and modulation.\cite{yan09.01} Yan et al. reported the fabrication of programmable architectures using array of nanowires.\cite{yan11.01}

Quantum Confined Structures (QCSs) for NLO applications were pioneered by Hache \emph{et al}, who showed the enhancement of the second hyperpolarizability of gold colloids to originate in the confinement of electrons and scale inversely proportional to the cube of the radius.\cite{hache86.02} Sanders and Chang studied the optical properties of Si quantum wires theoretically and showed that when the wire's widths is less that 8$\AA$, the excitonic oscillator strength is comparable with the one for bulk GaAs semiconductors.\cite{sande92.01} In other work, the effect of confinement on the luminescence of porous Si was emphasized.\cite{buda92.01} Chen \emph{et al.} showed that for Si quantum wires, the third order nonlinearity is proportional to $a^{-6}$, where $a$  is the Bohr radius of the exciton.\cite{chen93.01} Other studies also confirm the role of confinement in the NLO properties of quantum wires.\cite{xia97.01, balle99.01} The ability to control the confinement in a quantum system enables the control of NLO properties of these materials, leading to applications in optical devices including lasers, solar cells, and nonlinear-optical switches.\cite{banfi98.01}

Past studies have focused only  on exploiting the relation between confinement and nonlinear response of QCSs.  While the effect geometry on nonlinear response of organic materials has been the subject of several studies,\cite{ostro00.01, keina08.01, chu11.01, zhou11.01} no work has addressed this effect on artificial systems such as quantum graphs.

To the best of our knowledge, we investigate for the first time the role of geometry on NLO properties of quantum graphs, as 2-D networks of quantum wires, and show how the hyperpolarizability is affected by their shape and orientation.  Though past work also considered geometry,\cite{kuzyk06.02} these were for point charges with 2D Coulomb potentials, which are not easily reduced to practice as are quantum wires.  We show that geometry can lead to an enhancement in NLO response.  Our present focus is on quantum loops as a special class of quantum graphs; future work will include more complex graphs.

We believe we are also the first to investigate the nonlinear optical properties of quantum graphs and to develop a consistent, practical method for calculating quantum observables for graphs.  This includes establishing methodologies for computing matrix elements of arbitrary quantum operators (including double commutators leading to sum rules) between eigenstates of quantum graphs.  We freely use this lexicography in the text, referring the reader to the Appendices for a detailed description of the methodology.

Our paper is organized as follows.  First we separate geometric and confinement effects by imposing infinite transverse confinement to each wire, rendering it one-dimensional. Then, we use free-particle one electron wavefunctions and energies determined from the loop length to calculate the contributions to the nonlinear response of each segment. To test the viability of the results, we verify the 2D sum rules for the full loop. Finally, we employ numerical simulations to test the configuration space of geometries to find the largest hyperpolarizabilities. We conclude with a discussion of applications of this work.

\section{System of a single quantum wire}\label{sec:singleWire}
A quantum loop is composed of an arbitrary number of wire segments. The physics of the quantum graph reflects the contributions of single wires. For an extremely thin quantum wire, the electron moves freely along the wire but it is tightly confined in the transverse direction. This tight confinement eliminates size effects,\cite{shafe11.02} as we briefly show below.

The free-particle wavefunctions along the wire are
\begin{equation}\label{wireLongWavefunction}
\psi(s) = A e^{i k_s s} + B e^{- i k_s s},
\end{equation}
where $s$ denotes the longitudinal direction; $A$ and $B$ are constants; and $k_s$ is the longitudinal k-vector which is fixed via boundary conditions.

The uncertainty principle transverse to the wire under strong confinement implies infinite energy free particle states.\cite{hadji97} It can be shown \cite{shafe11.02} that these states do not contribute to the hyperpolarizability by modeling confinement using a highly weighted Delta potential:
\begin{equation}\label{transversePotential}
V(\tau) = -g \delta(\tau),
\end{equation}
where $\tau$ is the transverse coordinate and $g>>1$ is the potential strength. The transverse ground state of the electron has one bound state wavefunction given by
\begin{equation}\label{TransverseGround}
\eta_0 (\tau) = \sqrt{k_0^\tau}e ^ {- k_0^\tau \left| \tau \right| } ,
\end{equation}
where
\begin{equation}\label{definektau0}
k_0^\tau = \frac {m g}{\hbar^2},
\end{equation}
$m$ being the electron's mass with ground state energy of
\begin{equation}\label{transverseGroundEnergy}
E_0^\tau = - \frac {\hbar^2 {k_0^\tau}^2} {2 m}.
\end{equation}
The free state eigenfunctions in the limit of full confinement, i.e. $g \rightarrow \infty$, are given by
\begin{equation}\label{TransverseExcitedEven}
\eta_\nu^{even}(\tau) = -\sqrt{\frac{2}{a}}\cos(\nu\pi) \sin \left( \frac{2 \nu \pi }{a}|\tau| \right)
\end{equation}
and
\begin{equation}\label{TransverseExcitedOdd}
\eta_\nu^{odd}(\tau) = -\sqrt{\frac{2}{a}} \cos(\nu\pi) \sin \left( \frac{2 \nu \pi}{a} \tau \right),
\end{equation}
where $\nu$ takes integer values. The even and odd superscripts refer to the wavefunction's parity. Every single excited state is doubly degenerate with energy
\begin{equation}\label{transverseExcitedEnergy}
E_\nu = \frac {\hbar^2 {k_\nu^\tau}^2} {2m},
\end{equation}
where $k_\nu^{\tau}$ is found via
\begin{equation}\label{trans.Eq}
\frac {k_\nu^\tau a} {2} = \nu \pi.
\end{equation}
Here, $a$ is the size of the box transverse to the wire used for normalizing the free-particle wavefunctions.  When evaluating the contributions of the free-particle state, we take the infinite space limit $a\rightarrow \infty$.\cite{shafe11.02}

When $a>>1$ Eq. (\ref{transverseExcitedEnergy}) suggests that the transverse excited state energies are positive and become a continuum, and according to Eq. (\ref{transverseGroundEnergy}), the ground state energy approaches $-\infty$. Therefore, the energy gap becomes infinite so these states will not contribute to the nonlinear response.\cite{shafe11.02}

\section{Quantum loops}\label{sec:quantumLoop}

\begin{figure}
\includegraphics{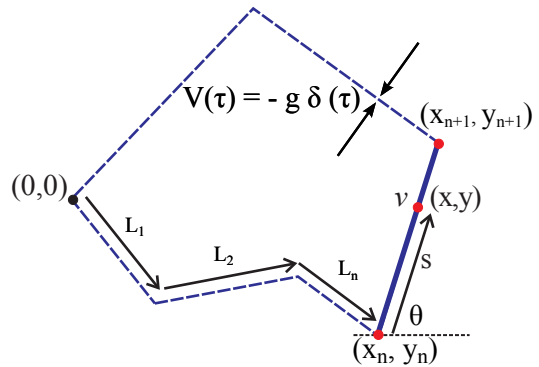}
\caption{A 2-D closed quantum loop. $(0,0)$ is the starting point of the loop. $(x, y)$ is the electron's coordinate at point $v$ with respect to $(x_n,y_n)$. $\theta$ is the angle that this wire segment makes with x axis. Each wire segment confines the electron in the transverse direction.}
\label{fig:WireLoop}
\end{figure}

The typical quantum loop shown in Fig. \ref{fig:WireLoop} is composed of N wire segments. The wavefunctions of an electron moving in this loop are free-particle states with periodic boundary conditions and flux conservation along the loop \cite{harri05.01}.  Latin sub or superscripts represent longitudinal and Greek characters represent the transverse directions. For example $E_{n}$ and $E_\nu$ denote the energy of the $n^{th}$ longitudinal and $\nu^{th}$ transverse states, respectively. Using this convention, the Hamiltonian of electron at each wire segment is given by
\begin{eqnarray}\label{totalH}
H(s,\tau) &=& -\frac{\hbar^2}{2 m}\left(\partial^2_s + \partial^2_\tau\right) - g\delta\left(\tau \right) \equiv H_s + H_\tau
\end{eqnarray}
where $\partial_x \equiv \partial/\partial x$ and
\begin{equation}\label{Hs}
H_s = -\frac{\hbar^2}{2 m}\partial^2_s \quad \mbox{and} \quad H_\tau = -\frac{\hbar^2}{2 m}\partial^2_\tau- g\delta\left(\tau \right).
\end{equation}
Using the separation of variables, the solution of the Schr\"{o}dinger Equation for the Hamiltonian in Eq. (\ref{totalH}) are
\begin{eqnarray}\label{longSegmentWavefunction}
\psi_l^i(s) &=& \frac {1} {\sqrt{R}} \exp \left[\pm\left(i k_l^s s + \phi_l^i \right)\right]\Theta \left[s\right]\Theta \left[L_i - s \right],
\end{eqnarray}
where $s$ is the electron's longitudinal coordinate measured from the starting point of the wire segment,
\begin{equation}\label{sDefine}
s = \sqrt{\left(x - x_{n}\right)^2+ \left(y - y_{n}\right)^2}.
\end{equation}
$R$ is the total length of the loop, $k_l^s$ is the longitudinal wave number given by,
\begin{equation}\label{ks}
k_l^s  = \frac{2 \pi l}{R},
\end{equation}
and
\begin{equation}\label{AccumulatedPhase}
\phi_l^i (s) = \frac{2 \pi l}{R} \cdot s(0,i-1),
\end{equation}
is the accumulated phase from the origin to the beginning of the wire segment and insures continuity of the wavefunction throughout the loop. $s_i(0,n)$ is the distance along the loop from the origin to the beginning of $i^{th}$ wire segment.
\begin{equation}\label{s(0,n)}
s(0,i-1) = \sum_{j=1}^{i-1} L_j = \sum_{j=1}^n \sqrt{\left(x_j - x_{j-1}\right)^2 + \left(y_j - y_{j-1}\right)^2}.
\end{equation}
The step-function in Eq. (\ref{longSegmentWavefunction}), defined as
\begin{equation}\label{stepFunction}
\Theta \left[x-x_0\right] = \Bigg\{\begin{array}{ll}
        1 & \quad \hbox{$x \geq x_0$} \\ \\
        0 & \quad \hbox{$x < x_0$}
      \end{array}
\end{equation}
demands that the wavefunction on $i^{th}$ segment vanishes everywhere except along the $i^{th}$ segment.

The longitudinal energy of the electron is found using periodic boundary conditions, $\psi(s+R) = \psi(s)$, yielding
\begin{equation}\label{longEnergyEigenvectors}
E_l^s  = \frac {2 \pi^2 \hbar^2 l^2} {m R^2} \qquad l = \pm 1, \pm 2, \cdots
\end{equation}
Given the symmetry between clockwise and counterclockwise motion, the longitudinal wavefunction is doubly degenerate.

The transverse wavefunctions of the electron at each wire segment are given by Eqs. (\ref{TransverseGround}), (\ref{TransverseExcitedEven}) and (\ref{TransverseExcitedOdd}).  The total wavefunction of the electron is the union of all segments including the transverse and longitudinal parts. Using the notation outlined in Appendix \ref{Sec:EdgeState}, the total wavefunction, $\phi_{m\mu}(s,\tau)$, can be written as a union over the graph of the wavefunctions $\phi_{m\mu}^{k}(s,\tau)$ on each edge:
\begin{equation}\label{WFexp}
\phi_{m\mu}(s,\tau)=\psi_{m}(s)\eta_{\mu}(\tau)=\cup_{k=1}^{E} \phi_{m\mu}^{k}(s,\tau)
\end{equation}

The total wavefunctions are orthogonal because they span a Hilbert space defined by their Hamiltonian.  However, the value of the total wavefunction on a particular segment has no quantum mechanical significance by itself:  All such contributions are required (with matched boundary conditions) to obtain the energy spectrum from a secular equation.  Consequently, the value of an eigenfucntion on a particular segment depends on its value on all other segments.  Two different modes on a particular segment don't have any special relationship between them.

Each energy eigenvalue is a sum of longitudinal and transverse energy components, given by Eqs. (\ref{longEnergyEigenvectors}) and (\ref{transverseGroundEnergy}) for the transverse ground state and (\ref{transverseExcitedEnergy}) for the transverse excited states, respectively.

The longitudinal excited state wavefunction are doubly degenerate because of the clockwise/counterclockwise symmetry of the Hamiltonian, and the transverse excited state wavefunctions are doubly degenerate and are eigenstates of the parity operator, with odd and even eigenvalues, in Eqs. (\ref{TransverseExcitedEven}) and (\ref{TransverseExcitedOdd}), respectively.

\section{Transition Moments}\label{sec:transitionMoments}
The hyperpolarizability is a function of energies and position matrix elements, $x_{ij}$'s, which we loosely call transition moments. The transition moments of an electron in a quantum loop are related to the sum over contributions from each segment by (see Eq. (\ref{WFexp}))
\begin{eqnarray}\label{xNM}
x_{p\kappa, q\lambda} &\equiv& \left< p\kappa|x|q\lambda \right> \\
&=& \int dxx \phi_{p\kappa}^{*}(x)\phi_{q\lambda}(x) \nonumber \\
&=& \sum_{k} \int_{i} dxx_{k} \ \phi_{p\kappa}^{k*}(x)\phi_{q\lambda}^{k}(x).\nonumber
\end{eqnarray}

For the $i^{th}$ wire segment, the $x$ coordinate of the electron with respect to the starting point of the graph, $(0,0)$ in Fig. \ref{fig:WireLoop}, is given by
\begin{equation}\label{xWireSegment}
x = s\cos\theta_i \pm \tau\sin\theta_i + x_1^i,
\end{equation}
where $\theta_i$ is the angle that segment $i$ makes with the $x$ axes, and $x_1^i$ is the $x$ coordinate of starting point of the segment $i$. `$\pm$' represents the two possible sides of the wire from which $\tau$ can be defined. The total transition moment is given by substituting for $x{_k}$ in Eq. (\ref{xNM}) and is the sum of a longitudinal contribution and a transverse contribution.

\subsubsection{Longitudinal Transition Moments}
For the two distinct cases given by $p\neq q$ and $p = q$, $\left(x_s^i\right)_{p\kappa, q\lambda}$, Eq. (\ref{PositionObserve})  yields,
\begin{description}
  \item[a] $p\neq q$;
\begin{eqnarray}\label{Longx1x2pq}
\left(x_s^i\right)_{p\kappa, q\lambda} &=& \frac{\delta_{\kappa\lambda}}{R}e^{ik_{qp}s(0,i-1)} \times \nonumber \\
&& \left[\left(\frac{\cos\theta_i }{k_{qp}^2}-\frac{x_2^i }{k_{qp}}i\right)e^{ik_{qp}L_i} - \frac{\cos\theta_i }{k_{qp}^2} +\frac{x_1^i}{k_{qp}}i\right]\nonumber \\
\end{eqnarray}
Note that $i=\sqrt{-1}$ should not be confused with the index $i$.
  \item[b] $p=q$;
\begin{eqnarray}\label{Longx1x2pp}
\left(x_s^i\right)_{p\kappa, p\lambda} &=& \delta_{\kappa\lambda}\frac{\left(x_{2}^i+x_{1}^i\right)}{2}\frac{L_i}{R}
\end{eqnarray}
\end{description}
Eqs. (\ref{Longx1x2pq}) and (\ref{Longx1x2pp}) suffice to calculate $x_{ij}$'s for any wire orientations including when $x_1 = x_2$ or $y_1=y_2$.

\subsubsection{Transverse Transition Moments}
We follow the same procedure to find the contributions from the transverse component. Due to parity selection rules, only odd (even) to even (odd) transitions are permitted. At each wire segment,
\begin{eqnarray}\label{xtau}
\left(x_{\tau}^{i}\right)_{p\kappa, q\lambda} &=& \sin\theta_i \int_{i} ds \psi_{p}^{i}(s) \psi_{q}^{i}(s) \int_{i} d\eta\eta \eta_{\kappa}(\tau) \eta_{\lambda}(\tau)\nonumber \\
&\equiv& \sin\theta_i\times I_{pq}^{i}\times J_{\kappa\lambda}^{i}
\end{eqnarray}
where
\begin{eqnarray}\label{sh-piqi}
I_{pq}^{i} = \left\{
      \begin{array}{ll}
        \frac{L_i}{R} & \hbox{$p=q$} \\ \\
        \frac{e^{i k_{qp}^s s(0,i-1)}}{i k_{qp}^s R} \left(e^{i k_{qp}^s L_i}-1\right) & \hbox{$p \neq q$}
      \end{array}
    \right.
\end{eqnarray}
and
\begin{eqnarray}\label{sh-kappaTauLambda}
J_{\kappa\lambda}^{i} = \quad \left\{
     \begin{array}{ll}
      \frac{-8\sqrt{2} (-1)^\kappa \pi \kappa a^{5/2} {k_0^\tau}^{3/2}}{\left(a^2 {k_0^\tau}^2 + 4 \pi^2 \kappa^2\right)^2} & \hbox{$ \kappa(\lambda) > 0, \lambda(\kappa) = 0$} \\ \\
       \frac{2(1+(-1)^{\kappa+\lambda + 1}) a \left|\kappa \lambda\right|} {\pi^2 \left(\kappa^2-\lambda^2\right)^2} & \hbox{$\lambda\neq \kappa\neq 0$} \\ \\
       \frac{a}{4} & \hbox{$ \left|\kappa\right|=\left|\lambda \right|\neq 0 $}
     \end{array}
   \right.
\end{eqnarray}

The graph's transition moments, Eq. (\ref{xWireSegment}), are obtained by summing over longitudinal and transverse transition moments from all wire segments, given by Eqs. (\ref{Longx1x2pq}) or (\ref{Longx1x2pp}), and (\ref{xtau}).

\section{Sum rules for quantum loops}\label{sec:SR for QL}
The Thomas-Reiche-Kuhn sum rules are a consequence of the commutators between the position operator and the Hamiltonian of the quantum system, $\left[x, \left[H, x\right]\right]$, and relate the transition moments and energies. In 1-D, the sum rules are given by
\begin{equation}\label{SR1D}
\sum_{n}\left(E_n - \frac{E_m + E_p}{2}\right)x_{mn}x_{np} =
\frac{\hbar^2}{2m}\delta_{mp}
\end{equation}
In general, the summation spans the complete set of eigenstates of the system, including both discrete and continuum \cite{Bethe77.01} and degenerate and non-degenerate states \cite{ferna02.01}.

Sum rules have been widely applied.\cite{bello08.01} In nonlinear optics, they are used to calculate the fundamental limits of first and second hyperpolarizabilities,\cite{kuzyk00.01, kuzyk00.02} to find a dipole-free sum over states expression for calculating the nonlinear hyperpolarizabilities\cite{kuzyk05.02, perez01.08}, and to impose constraints on the related parameters in Monte Carlo simulations of the first\cite{kuzyk08.01} and second\cite{shafe10.01} hyperpolarizabilities.  In this section we obtain an analytical relation for sum rules in the longitudinal and transverse directions of a quantum graph and use these to validate the calculated eigenfunctions.

Sum rules in 3-D were verified for the quantum rigid rotators by Hadjimichael \emph{et al.}, who showed that the classical picture of a rigid rotator, where the radius of rotation is fixed, is consistent with the sum rules when taking into account the radial component of the wavefunction.\cite{hadji97} They used the uncertainty principle in the radial component under a highly-confining delta function potential, $V(r) = -g \delta(r-R_0)$ with $g > 0$ and $R_0$ being the potential strength and radius of the rotator, respectively, to show that the continuum states corresponding to high radial momentum ensure that the sum rules are obeyed.

In reference \cite{shafe11.02}, it was shown that Eq. (\ref{SR1D}) for a single quantum wire may be written as a sum rule over $s$ plus a sum rule over $\tau$.  Since the sum rule is derived from a double commutator of x with H, it is an operator which may be evaluated across the graph using Eq. (\ref{sumruleInEdges}) in Appendix \ref{Sec:EdgeState} as a sum over edges of the sum rules for each wire segment:
\begin{eqnarray}\label{quantumLoopSumRules}
&&\delta_{\kappa\lambda} \sum_{i,n} \left[ \left( E_n^s - \frac{E_p^s + E_q^s}{2}\right)x_{pn}^{s,i} x_{nq}^{s,i} \right]\nonumber \\
&+& \sum_{i,\nu} \left[\left(E_\nu^\tau - \frac{E_\kappa^\tau + E_\lambda^\tau}{2}\right) I_{pq}^{i} x_{\kappa\nu}^{\tau,i} x_{\nu\lambda}^{\tau,i}\right] \nonumber \\
&=& \frac{\hbar^2}{2m}\delta_{pq}\delta_{\kappa\lambda}.
\end{eqnarray}
where $I_{pq}^{i}$ is the edge overlap integral defined in Eqn (\ref{sh-piqi}).  In deriving Eq. (\ref{quantumLoopSumRules}), we used the closure relation Eq. (\ref{outer}) to write $\sum_{n,j} I_{pn}^{i}I_{nq}^{j} = I_{pq}^{i}$.

The first and second terms on the left-hand side of Eq. (\ref{quantumLoopSumRules}) can be interpreted as the longitudinal and transverse contributions, respectively.

\begin{table}
\caption{Numerical results for diagonal components, i.e. $p=q$ and $\kappa = \lambda$ in Eq. (\ref{quantumLoopSumRules}) for triangles, $100$ longitudinal and $200$ transverse states.}\label{table:dataSet1}
\begin{tabular}{|l ||l c l c|}
  \hline
  Triangle Coordinates &  & Longitudinal &  & Transverse \\
  $(x_0, y_0, x_1, y_1, x_2, y_2)$ &  & ($\times \hbar^2/m$) &  & ($\times \hbar^2/m$). \\ \hline \hline
  (23,-16,3,-19,24,-15) &  & 0.474 &  & 0.022 \\
  (18,-7,21,1,-25,-20) &  & 0.401 &  & 0.095\\
  (-32,-21,-49,-42,-69,-88) &  & 0.118 &  & 0.378 \\
  (3,-93,-13,36,35,20) &  & 0.093 &  & 0.403 \\
  (-7,-16,-27,64,42,-23) &  & 0.196 &  & 0.299 \\
  \hline
\end{tabular}
\end{table}
We can use Eq. (\ref{quantumLoopSumRules}) to test the validity of the results in section \ref{sec:quantumLoop}. We pick the simplest possible loops, i.e. triangles, made of three quantum wires connected at vertices. Many configurations are tested by randomly sampling the vertices.  The energy and transition moments for each are calculated as previously described, and evaluated by the expression on the right-hand side of Eq. (\ref{quantumLoopSumRules}). The numerical result for six different triangles are presented in table \ref{table:dataSet1} for diagonal sum rules only.  Verifying non-diagonal sum rules follows the same procedure. The sum of second and third columns in Table \ref{table:dataSet1}, when multiplied by $\hbar^2/m$ yields the left-hand side of Eq. (\ref{quantumLoopSumRules}), thus confirming the validity of our protocol.

The projections of the longitudinal transition moments on the lab coordinate axes depend on the \emph{geometry} through the orientations of the wire segments. Consequently, the nonlinear optical response of quantum loop depends on the shape and its orientation.  In contrast, the projections of the transition moments due to the transverse and longitudinal coordinates together leave the sum rules invariant to the loop's shape or orientation.

The transverse wavefunctions do not contribute to $\beta$ and $\gamma$ because of the infinite energy denominators. However, they make finite contributions to sum rules because the product of infinitesimal transition moments and infinite energy $E_{ij} x_{ij}$ is finite.  Hence, when calculating the hyperpolarizability we only use the longitudinal parts given by Eqs. (\ref{Longx1x2pq}) and (\ref{Longx1x2pp}).

\section{Geometry-Controlled Nonlinear Response}\label{sec:NLO}

\subsection{First Hyperpolarizability}

In this section, we investigate the effect of geometry on the hyperpolarizability of a quantum graph. The off-resonance diagonal component of the hyperpolarizability, $\beta_{xxx}$, is given by \cite{orr71.01},
\begin{equation}\label{beta}
\beta_{xxx} = -3e^3 {\sum_{n,m}}' \frac{x_{0n}\bar{x}_{nm}x_{m0}}{E_{n0}E_{m0}},
\end{equation}
where $x_{ij}$ is the transition moment of the quantum graph along the coordinate axis $x$. $\bar{x}_{nm} = x_{nm}-x_{00} \delta_{nm}$ and the prime in the summation indicates that the ground state is excluded from the summation.  The fundamental limit formalism defines $\beta$ to be the {\em largest} diagonal tensor element. For a given quantum system, $\beta$ is determined by finding the coordinate frame in which $\beta_{xxx}$ is the largest.  In the present work, we do so by fixing the coordinate axes and determining $\beta_{xxx}$ as a function of angle, from which we determine the peak value $\beta$.  One can take the same approach to calculate the effects of geometry on other components such as $\beta_{xxy}$ and $\beta_{xyy}$.

For a quantum system with $N$ electrons and energy difference between first excited state and ground state, $E_{10}$, the theory of FL predicts that the largest attainable diagonal element of the hyperpolarizability, $\beta_{max}$, is given by \cite{kuzyk00.01}
\begin{equation}\label{FL}
\beta_{max} = 3^{1/4} \left(\frac{e \hbar}{m^{1/2}}\right)^{3}\frac{N^{3/2}}{E_{10}^{7/2}}.
\end{equation}
To optimize the nonlinear response regardless of the molecular size, we use the \emph{intrinsic} hyperpolarizability, $\beta_{int}$, as a scale-invariant figure of merit,\cite{kuzyk10.01}
\begin{eqnarray}\label{betaInt}
\beta_{int} &\equiv& \frac{\beta}{\beta_{max}} = \left(\frac{3}{4}\right)^{3/4} {\sum_{n,m}}' \frac{\xi_{0n}\xi_{nm}\xi_{m0}}{e_n e_m} ,
\end{eqnarray}
where $\xi_{ij}$ and $e_i$ are normalized transition moments and energies, defined by
\begin{equation}\label{normalizedTrnsition}
\xi_{ij} = \frac{x_{ij}}{x_{01}^{max}}, \qquad e_{i} = \frac{E_{i0}}{E_{10}},
\end{equation}
where
\begin{equation}\label{Xmax}
x_{01}^{max} = \left(\frac{\hbar^2}{2 m E_{10}}\right)^{1/2} ,
\end{equation}
and where $x_{01}^{max}$ represents the largest possible transition moment, $x_{01}$.  According to Eq. (\ref{normalizedTrnsition}), $e_0 = 0$ and $e_1 = 1$. $\beta_{int}$ is scale-invariant and can be used to compare molecules of different shapes and sizes.

We sample triangle configurations by randomly choosing the vertex coordinates, then calculating the total energy and transition moments using Eqs. (\ref{longEnergyEigenvectors}), (\ref{Longx1x2pp}) and (\ref{Longx1x2pq}). Subsequently, $\beta_{int}$ is calculated using Eqs. (\ref{betaInt}) and (\ref{normalizedTrnsition}). For each configuration, the triangle is rotated in x-y plane in increments of $15^{\circ}$ and $\beta_{int}$ is calculated for each orientation. The largest $\beta_{int}$ value and the corresponding vertex coordinates are recorded.  After thousands of trials, the best intrinsic hyperpolarizability found is $ |\beta_{int}| = 0.049$.

\begin{figure}
\includegraphics{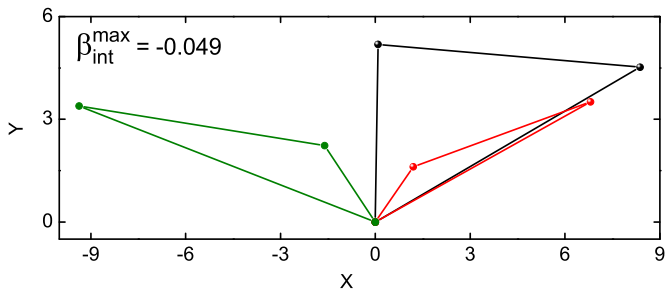}\\
\caption{The three triangles with largest $\beta_{int}(\simeq -0.049)$.}\label{fig:triangle}
\end{figure}

Figure \ref{fig:triangle} shows the three triangles with largest $\beta_{int} \simeq -0.049$. Though the triangles differ in area, the scale invariance of $\beta_{int}$ assures that they can be compared.  Rotating each triangle by $180^\circ$ identifies the orientation with peak $\beta_{int}$ of positive sign.

\begin{figure}
\includegraphics{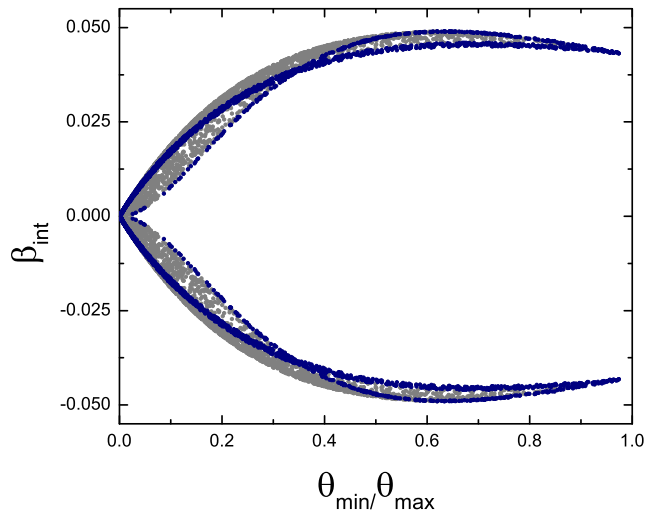}\\
\caption{Distribution of $\beta_{int}$ over its full range of triangular configurations as quantified by the ratio of the smallest angle, $\theta_{min}$, and largest angle, $\theta_{max}$. Points in blue represent isosceles triangles.}\label{fig:betaAngles}
\end{figure}

Figure \ref{fig:betaAngles} plots $\beta_{xxx}$ versus the ratio of the smallest to the largest internal angles,  $\theta_{min}/ \theta_{max}$ - a measure of the shape of a triangle. The blue points label the isosceles triangles.

\begin{figure}
  \includegraphics{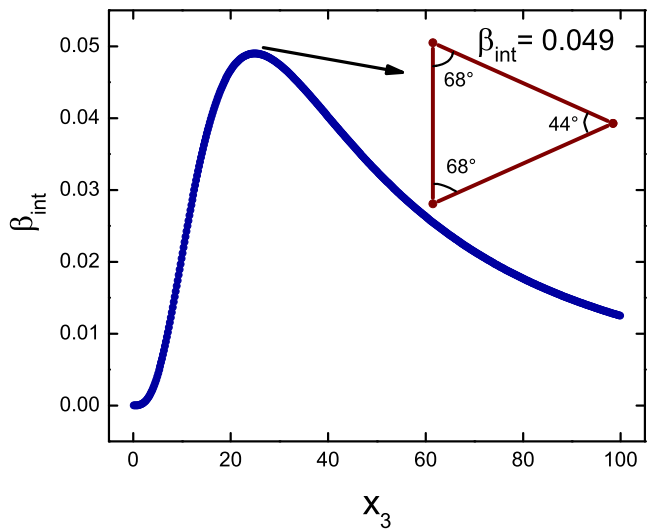}\\
  \caption{$\beta_{int}$ for isosceles triangles with two points fixed at (0,10) and (0,-10), and the third point varying along the x axis. The triangle in the picture represents the largest value of $\beta_{int} = 0.049$}\label{fig:isosceles}
\end{figure}

Since simulations show isosceles triangles with the largest hyperpolarizabilities, we more systematically study them by fixing two of the vertices at coordinates $(0,10)$ and $(0,-10)$ while the third vertex is scanned along $y=0$  for $x_3>0$. Fig. \ref{fig:isosceles} shows a plot of $\beta_{int}$ as a function of $x_3$ yielding a peak of $\beta_{int} \approx 0.049$ for the triangle in the inset.  This peak value is the same for all triangles with two angles $ \simeq 68^\circ$ and $\simeq 44^\circ$, independent of their size.  Previous studies that varied the positions and charges of 4 point nuclei found that the optimum geometry with $72^0$ and $35^o$ yielded $\beta_{int} \approx 0.65$.\cite{kuzyk06.02}  However, it is not possible to control point nuclei with such precision, let alone synthesis molecules with arbitrarily charged nuclei and bond angles.  This is why real molecules fall far short of the limits.

Studies which optimize the shape of the potential energy function yield the largest possible nonlinear optical response of $\beta_{int} \simeq 0.71$ for a large set of potentials, which universally have the property that $X = x_{01}/x_{01}^{max}$ where $x_{01}^{max}$ is given by Eq. (\ref{normalizedTrnsition}).\cite{zhou06.01,zhou07.02, ather12.01} Since the optimized triangle has $X = 0.49$, isosceles triangles are not globally optimized, but represent a local maximum under constraints of their geometry. In future studies, we will consider other graphs, including bent wires, three-pronged star graphs and fractals and will show that a much larger value of $\beta_{int}$ possible.

In the three-level ansatz, $\beta_{int}$ can be expressed as,\cite{kuzyk09.01}
\begin{equation}\label{betaIntfG}
\beta_{int} = f(E)G(X) ,
\end{equation}
where
\begin{equation}\label{f(E)}
f(E) = (1-E)^{3/2} \left( E^2 + \frac {3} {2} E + 1 \right),
\end{equation}
\begin{equation}\label{G(X)}
G(X) = \sqrt[4]{3} X \sqrt{\frac {3} {2} \left( 1 - X^4\right)},
\end{equation}
and $E=E_{10}/E_{20}$. The energy spectrum for a triangle is similar to that of a particle in a box, which has large intrinsic hyperpolarizability.\cite{shafe11.01} For a triangle with $E = 3/8$ and $X = 0.49$, in principle, the hyperpolarizability could be as high as $\beta_{int} = 0.65$ for a three-level system. Comparing this value with $\beta_{int} \simeq 0.05$ suggests that a larger number of states contribute to the nonlinear optical response. Also, FL theory predicts that degeneracy lowers the response. $\beta_{ij}^{int}$ for the optimized triangle shows that 5 states contribute significantly.  These two factors suppress the full potential of a triangle.

The search for best NLO quantum loops can be generalized to 3+ segments. As for triangles, the coordinates of four vertices are chosen randomly without restrictions. In configurations where wire segments cross, we assume that there is no transfer of probability current between the two -- for example one segment can be viewed as being above the other one by an infinitesimal distance with tunneling effects neglected. The best quadrilateral gives $\beta_{int} = 0.073$, which is 50\% larger than for the best triangles. For a larger number of segments, the improvement in $\beta_{int}$ is marginal. Future work will address more complex graphs including many-electron band-filling models.

\subsection{Second Hyperpolarizability}

The sampling approach for calculating $\beta$ for loops may be applied to the calculation of the second hyperpolarizability, $\gamma$, using the same matrix elements that we used for $\beta$.  In these studies, we seek to optimize the intrinsic second hyperpolarizability given by
\begin{eqnarray}\label{gammaInt}
\gamma_{int} &=& \frac{1}{4} \left({\sum_{n,m,l}}' \frac{\xi_{0n} \bar{\xi}_{nm}\bar{\xi}_{ml}\xi_{l0}}{e_n e_m e_l} - {\sum_{n,m}}' \frac{|\xi_{0n}|^2 |\xi_{m0}|^2}{e_n^2 e_m}\right) , \nonumber \\
\end{eqnarray}
where the normalized transition moments, $\xi_{ij}$, and energies, $e_i$, were introduced in Eq. \ref{normalizedTrnsition}.

\begin{figure}
\includegraphics{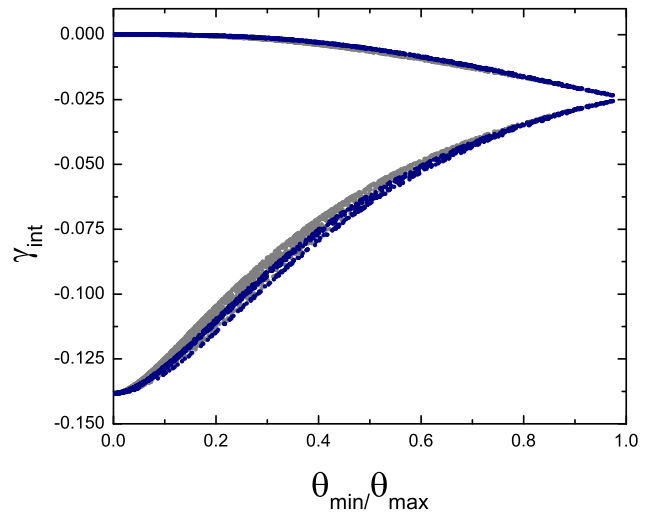}\\
\caption{Distribution of $\gamma_{int}$ over its full range of triangular configurations as quantified by the ratio of the smallest to largest angle.}\label{fig:gammaAngles}
\end{figure}

The result for a triangle loop yields a maximum value $\gamma_{xxxx}=0$ for high aspect ratio (flat or squashed) triangles lying mainly along the y-axis and the largest negative value of -0.138 for similarly-shaped triangles lying nearly parallel to the x-axis.  Fig \ref{fig:gammaAngles} displays the variation of $\gamma_{xxxx}$ over the entire range of aspect ratios for all 10,000 triangles sampled in the Monte Carlo run.

The two branches correspond to the minimum and maximum values of the intrinsic hyperpolarizability for a given aspect ratio as quantified by $\theta_{min} / \theta_{max}$.  For an  equilateral triangle, $\theta_{min} / \theta_{max} = 1$, the two branches merge to yield the same minimum and maximum values simultaneous.  At the other extremum of a large aspect ratio triangle, the minimum and maximum values of $\gamma_{xxxx}$ are at the extremes, suggesting that large aspect ratio leads to the largest magnitude of the second hyperpolarizability.

Similar results hold for loops with four edges.  The second hyperpolarizability is always negative, and its minimum is at about -0.138.  This remarkable result illustrates that graphs with identical topologies but different geometries have similar nonlinear optical responses.  This is the topic of a future paper.

Comparison between Figures \ref{fig:betaAngles} and \ref{fig:gammaAngles} reveals that there may be a relationship between the two.  In both branches, the magnitude of $\beta_{int}$ increases as the triangle becomes open.  In comparison, the largest magnitude of $\gamma_{int}$ decreases as the triangle becomes more open.  Thus, there may be a complementarity between the two, as is often found in other properties.

\section{Concluding Remarks}
Low dimensional quantum systems are scientifically interesting because of their novel quantum mechanical effects, and should have a broad range of potential applications in future technologies. The fact that the optical properties of these systems can be manipulated by varying the degree of confinement provides a useful tool for the fabrication of novel nano-scale structures for new device concepts and to provide novel approaches for making more efficient nonlinear materials.  The present work has established a theoretical foundation for using quantum graphs as prototypes for investigating how geometry impacts the NLO response.

We deliberately prepared a single quantum wire in a way that made confinement effects irrelevant to the nonlinear response. Then we solved the Schr\"{o}dinger Equation for an electron in a wire as part of a larger network. The 2-D sum rules were obtained and were used to test the validity of the results. It was shown that the transverse wavefunction of the electron is essential in validating sum rules, but plays no role in the NLO response. Thus the hyperpolarizability of the quantum graphs with different shapes and identical topology are solely controlled by geometrical effects.

Monte Carlo simulations identified the best triangle loops with the largest $\beta_{int}$ values as a unique class of isosceles triangles with orientations and internal angles given in Fig. \ref{fig:isosceles}. 4-segment loops show even larger values.  Preliminary results for rectangular loops show an intrinsic value near 0.07. While this value is far from the fundamental limit, this geometry-controlled value is better than the hyperpolarizability of all other known nanosystems.  This study, which focuses on geometrical determinants of $\beta_{int}$ and $\gamma_{int}$ shows promising results toward replacing nonlinear optical materials with artificially structured systems. Recent work on more complex graphs have yielded much higher intrinsic hyperpolarizabilities.\cite{lytel12.01}

We also showed that the triangle graphs have negative $\gamma_{int}$.  This appears to be a consequence of the closed-loop topology of the graph and not its geometry.  Parallel studies underway in our group have shown that opening a vertex of the triangle without changing its geometry allows for positive values of $\gamma_{xxxx}$.

The next potential direction in exploring the nonlinear optical response of quantum graphs would be generalizing the one-electron model to N-electrons, and considering the resonant regime, where the theory can be compared with experimental results and be more relevant for practical applications. Meanwhile, it is also interesting to study more complex graphs, from the more exotic three-pronged star graph \cite{pasto09.01} to fractal systems, where the extra degrees of freedom lead to greater design flexibility.

Additionally, the mathematical formalism that we have developed will be of value in evaluating more complex graphs by relating the properties of individual nanowires to the properties of complex structures in which they are fundamental building blocks.

\begin{acknowledgments}
S. Shafei and M. G. Kuzyk thank the National Science Foundation (NSF) (ECCS-1128076) and Wright Paterson Air Force Base for generously supporting this work.
\end{acknowledgments}

\appendix
\section{Quantum graph lexicography}\label{Sec:EdgeState}
A quantum graph is a metric graph supporting particle dynamics. It is characterized by a Hamiltonian $H$ with a complete set of eigenstates $|N\rangle$ and an energy spectrum $E_{N}$ satisfying $H|N\rangle = E_{N}|N\rangle$.  If the graph is planar and the particle is confined to move on the $E$ edges, the graph is a quasi 1-D system.  We fix the graph in the xy-plane.

The energy eigenstates $|N\rangle$ span a Hilbert space and are constructed from a union of non-intersecting subspaces, each spanned by vectors $|N^{k})$, where $k$ labels an edge, which also satisfy $H|N^{k}) = E_{N}|N^{k})$.  The eigenstates $\left| N \right>$ may be written as a union of edge states with the interpretation that the dynamics of the particle on edge $k$ is described by the set of edge states $|N^{k})$ whose projections into real space $\langle x|N^{k})\equiv\phi_{N}^{k}(x)$ are wavefunctions exactly equal to the projection of the eigenstates $\psi_{N}(x)\equiv\langle x|N\rangle$ on edge $k$.  The eigenstates and operators acting on them comprise the Hilbert space of the graph.

Graph dynamics, including nonlinear optical phenomena, are calculated from the eigenstates.  Quantum mechanical relations, such as matrix elements or sum rules, are expressed in terms of eigenstates, not edge states.  Edge states are not complete on the graph:  A set of edge states on the $k^{th}$ edge represents dynamics on that specific edge, not on other edges. This implies that inner products of eigenstates will necessarily be sums of inner products of edge states from identical edges.  In real space, this simply means that the integral over the graph of a bilinear form of eigenstates with any operator is parsed into a sum of integrals over edges of bilinear forms of edge wavefunctions.  Bilinear forms such as $\left(N^{i}|M^{j}\right)$ are meaningless unless $i=j$.

The edge wavefunctions form a complete set on an edge only when every eigenfunction on an edge obeys the same boundary conditions at the ends of the edge.  Eigenfunctions that do not obey such boundary conditions are not complete on an edge, but, the wavefunction composed of the union of all edges produces a complete set of eigenfunctions of the Hamiltonian on the full graph.  The literature often refers to edge spaces as Hilbert spaces, though they are actually subspaces that only in union produce a Hilbert space representing a dynamical quantum system.

In position space, solving a graph means finding the edge wavefunctions and energy spectrum, matching boundary conditions at the vertices, and constructing the eigenfunctions as a union of edge wavefunctions using an appropriate notation.  This approach ensures that the abstract definition of the quantum graph is realized properly in position space.  All physical quantities may be calculated using the edge wavefunctions and the energy spectrum, provided that the definition of the union operation is properly implemented.

To this end, it is useful to define the union operation in abstract space and use it to derive expressions for matrix elements of operators, inner and outer products of eigenstates, and sum rules. All of these may be reduced to sums over the graph of integrals of bilinear products of edge wavefunctions and real-space operators.  But the abstract form enables almost instant proof of theorems such as sum rules for graphs for which proof has been shown for single-edged graphs (quantum wires).

We emphasize here that the following abstract definitions are exactly that -- abstractions defined to facilitate the manipulation of edge states comprising true eigenstates when physical properties are to be calculated.  The abstraction is merely book-keeping; the definitions could easily well be written using alternate notation.

We define an energy eigenstate vector as
\begin{equation}\label{ket}
\left| N \right> = \cup_{k=1}^E \left| N^k \right)
\end{equation}
and the dual (bra) state as
\begin{equation}\label{bra}
\left< M \right| = \left( M^j \right| \cup_{j=1}^{\dag E} ,
\end{equation}
where the union is over all the edges, $E$.  The operation $\cup_{k=1}^E$ is placed to the left of the ket while its conjugate is placed to the right.  We do this for convenience and as a reminder of what gets 'unioned' to what.  The union operation is not an operator in any sense ordinarily used in Hilbert space; it is an instruction that can be placed anywhere that is convenient, and can be interpreted accordingly.

The union operator \emph{always} accompanies an edge bra or ket in any expression involving more than one edge of the graph.  Note that the Hamiltonian operating on an energy eigenvector results in $H\left| N \right> = \cup_{k=1}^E H\left| N^k \right)=E_{N}\cup_{k=1}^E \left| N^k \right)=E_{N}\left| N \right>$ since all edges have the same energy spectrum as the graph.

The inner product of two energy eigenstates in this notation is then
\begin{equation}\label{inner}
\langle N|M\rangle = \left( M^j \right| \cup_{j=1}^{\dag E} \cup_{i=1}^E \left| N^i \right) =\delta_{NM} \\
\end{equation}
since the energy eigenstates are orthonormal.  The inner product is defined as:
\begin{eqnarray}\label{inner2}
\langle N|M\rangle &=& \int ds \, \psi^* (s) \psi(s) = \sum_k^E \int dx_k \, \phi_N^{k*} (x_k) \phi_M^k (x_k) \nonumber \\
&=& \sum_{k=1}^{E} \left( M^k | N^k \right) .
\end{eqnarray}
Equations \ref{inner} and \ref{inner2} define the allowed operations from which all others follow.

Similarly, we may compute the outer product of two eigenstates as
\begin{equation}\label{outer}
|N\rangle\langle M| = \cup_{j=1}^{E}|M^j)(N^i|\cup_{i=1}^{\dag E }.
\end{equation}
When Eq. (\ref{outer}) is inserted between two operators or eigenstates, the union operations act to select out the proper edges on either side of the outer product to ensure that all physical quantities are calculated on the same edge, then summed over edges.

The definitions in Eqs. (\ref{ket}) and (\ref{bra}) may now be used to express the matrix elements of an operator as sums over edge states in a way similar to the derivation of Equation \ref{inner2}:
\begin{eqnarray}\label{ME}
\langle N|O|M\rangle &=& \left( N^j \right| \cup_{j=1}^{\dag E} O \cup_{i=1}^E \left| M^i \right)\nonumber \\
&=& \sum_{i=1}^{E} \left( N^i |O| M^i \right)
\end{eqnarray}

The completeness relationship in Eq. (\ref{outer}) may be used with the definitions above to derive the matrix element of a composite operator AB by inserting it between AB.  It is equivalent to the standard approach of inserting a complete set of eigenstates between AB and using the definitions of the bra and ket expansions in edge states to find:
\begin{eqnarray}\label{ME2}
&&\langle N|AB|M\rangle \\
&=& \sum_{P} \langle N|A|P\rangle\langle P|B|M\rangle\nonumber \\
&=& \sum_{P} \left( N^j \right| \cup_{j=1}^{\dag E} A \cup_{i=1}^E \left| P^i \right)\left( P^k \right| \cup_{k=1}^{\dag E} B \cup_{l=1}^E \left| M^l \right)\nonumber \\
&=& \sum_{P} \left(\sum_{j=1}^E (N^{j}|A|P^{j})\right) \left(\sum_{l=1}^E (P^{l}|B|M^{l})\right)\nonumber
\end{eqnarray}
where the last step follows from Eq. (\ref{inner}) and the fact that the unions do not act upon the operators themselves.  This approach to expressing matrix elements of composite operators in sums over edge state matrix elements may be scaled repeatedly to any number of operators.

To compare the use of eigenstates and edge states, consider the derivation of the the sum rules.  The sum rules follow from calculating the matrix elements of the commutator $[[H,x],x]$,\cite{kuzyk01.01}
\begin{eqnarray}\label{sumrule}
\langle N|\left[[H,x],x\right]|M\rangle &=& \langle N|Hxx+xxH-2xHx|M\rangle \nonumber \\
&=& \frac {\hbar^2} {2m} \delta_{NM}.
\end{eqnarray}
Using closure, the first term may be written as
\begin{eqnarray}\label{HxxTerm}
\langle N|Hxx|M\rangle &=& \sum_{P} E_{N}\langle N|x|P\rangle\langle P|x|M\rangle. \nonumber \\
\end{eqnarray}
Expressing the other terms in this same way and summing them leads to the usual TRK sum rules.\cite{kuzyk01.01}.

It is now a simple exercise to express the TRK sum rules in terms of edge states by using Eq. (\ref{ME}) with $O=[[H,x],x]$.  By using the closure relationship Eq. (\ref{outer}) between pairs of operators, the result is that
\begin{eqnarray}\label{sumruleInEdges}
&&\sum_{i,j=1}^{E} \left(\sum_{P} [E_{P}-\frac{1}{2}(E_{N}+E_{M})](N^{i}|x|P^{i})(P^{j}|x|M^{j})\right) \nonumber \\
&=&\frac {\hbar^2} {2m} \sum_{i=1}^{E} (N^{i}|M^{i}) \\
&=&\frac {\hbar^2} {2m} \delta_{NM}\nonumber
\end{eqnarray}

It has been shown that the TRK sum rule decomposes into a longitudinal and transverse part on a single wire (edge) \cite{shafe11.02} by using the commutator of H with both longitudinal and transverse coordinates for the wire.  The general term on the left hand side of Eq. (\ref{sumruleInEdges}) may be written as the product of matrix elements between edge states of the double commutator, since edge states are themselves eigenvectors of H. Thus, the decomposition of a single wire sum rule into longitudinal and transverse parts holds exactly for the sum over edges of all such wires.

This formalism can be generalized to more complex graphs and curved edges by either converting sums to integrals in the infinitesimal edge limit, or by defining a curvature metric that maps an edge to a curve.

A final remark about edge functions:  They must not be used on their own to compute global quantities for a graph, but must always appear in unions over edges.  However, it is likely that graphs with similar motifs embedded in their topology will have edges with similar wavefunctions.  As an example, a graph with two loops separated by a single wire (a so-called barbell graph) has edge wavefunctions on the loops that are identical in form and differ only in details such as edge lengths.  When writing a set of solutions from scratch for a new graph, one can write down the form of the edge functions almost immediately and even select their amplitudes so they match easily to the connecting wire.  The specific relations among the amplitudes for the loops and connecting wire are then found from the conservation of flux equations, which also lead to the energy spectrum for the graph.  It is unknown at this time whether or not there are sum rules associated with individual edges.

\section{Calculation of transition moments}

As represented in the above formalism, the absolute position and orientation of an edge relative to the lab frame is not explicitly labeled in an edge ket in E-space. In the an edge's frame, we define $\hat{s}$ along the edge and $\hat{\tau}$ perpendicular to $\hat{s}$ (i.e. Eq. (\ref{WFexp}) with $x^f \rightarrow s$ and $y^f \rightarrow \tau$).  We limit ourselves to two dimensions, though edges in higher dimensions are treatable in the same way.  The wavefunction along an edge $k$ is $\phi_k (s) = \left<s \right| \left. \phi^k \right)$.  The same holds for $\tau$.  We may choose either end as the starting point.

When calculating the expectation value of an observable in the lab frame, we can transform the wave function to the lab frame, or we can transform the observable into the edge frame.  We choose the latter.  For edge $k$ with the reference end at $x_1^k$ and $y_1^k$, the position operator can be expressed as,
\begin{eqnarray}\label{PositionX}
x^k &=& \mathbb{1} x_1^k + \mathbb{1}_{\tau} s \cos \theta  - \mathbb{1}_{s} \tau \sin \theta \nonumber \\
y^k &=& \mathbb{1} x_1^k + \mathbb{1}_{\tau} s \sin \theta  + \mathbb{1}_{s} \tau \cos \theta  ,
\end{eqnarray}
where $\mathbb{1}_{s}$ and $\mathbb{1}_{\tau}$ are the identity operators in the subspace $s$ and $\tau$.

The contribution from edge $k$ to the position matrix element is given by our conventions as defined above,
\begin{eqnarray}\label{PositionObserve}
\left(p \kappa \right| x^k \left| q \lambda \right) &=& \delta_{\kappa,\lambda} \int_0^{L^k} ds \, \psi_p^{k*} (s) \left(  x_1^k + s \cos \theta \right) \psi_q^k (s) \nonumber \\
&-& I_{pq} \int_{-\infty}^{+\infty} \cancelto{0}{ d \tau \, \phi_p^{k*} (\tau) \tau \sin \theta \phi_p^k (\tau)} ,
\end{eqnarray}
where $p$ and $q$ are quantum numbers for states along the edge; $\kappa$ and $\lambda$ are quantum numbers for the transverse states, and $L^k$ is the length of edge $k$.  The edge overlap integral $I^{k}_{pq}$ was defined in Eq. (\ref{xtau}). The second term was shown to vanish for tight transverse confinement.\cite{shafe11.02} This illustrates the derivations of Eqs. (\ref{Longx1x2pq}) and (\ref{Longx1x2pp}), with the definition $\left(x_s^i\right)_{p\kappa, q\lambda} = \left(p \kappa \right| x^k \left| q \lambda \right)$ and a modicum of algebra.


\end{document}